\documentclass[intlimits,twoside,a4paper]{article}
\usepackage{amsmath,amssymb}
\usepackage{graphicx}

\usepackage[T2A]{fontenc}
\usepackage[cp1251]{inputenc}

\usepackage[eqsecnum]{cmpj2}

\issue{2015}{18}{1}{13606}
\doinumber{10.5488/CMP.18.13606}

\title[Heat capacity of liquids]%
{Heat capacity of liquids: A hydrodynamic approach}%

\author[T. Bryk, T. Scopigno, G. Ruocco]{T.~Bryk\refaddr{label1,label2}, T.~Scopigno\refaddr{label3}, G.~Ruocco\refaddr{label3,label4}}

\authorcopyright{T. Bryk, T. Scopigno, G. Ruocco, 2015}

\addresses{
\addr{label1}Institute for Condensed Matter Physics of the National Academy of Sciences of Ukraine,\\
1 Svientsitskii St., 79011 Lviv, Ukraine
\addr{label2} Institute of Applied Mathematics and
Fundamental Sciences, Lviv Polytechnic National University, \\79013
Lviv, Ukraine
\addr{label3} Dipartimento di Fisica, Universita di
Roma La Sapienza, I--00185, Roma, Italy
\addr{label4} Center for Life Nano Science @Sapienza, Istituto Italiano di
  Tecnologia, \\295 Viale Regina Elena, I--00161, Roma, Italy
}

\date{Received March 3, 2015}

\begin{document}
\maketitle

\begin{abstract}
We study autocorrelation functions of energy, heat
and entropy densities obtained by molecular dynamics simulations of supercritical Ar
and compare them with the predictions of the hydrodynamic theory.
It is shown that the predicted by the hydrodynamic theory single-exponential shape
of the entropy density autocorrelation functions is perfectly reproduced for small
wave numbers by the molecular dynamics simulations and permits the calculation of the
wavenumber-dependent specific heat at constant pressure.
The estimated wavenumber-dependent specific heats at constant volume and pressure,
$C_{\mathrm{v}}(k)$ and $C_{\mathrm{p}}(k)$, are shown to be in the long-wavelength limit in good agreement
with the macroscopic experimental values of $C_{\mathrm{v}}$ and $C_{\mathrm{p}}$ for the studied
thermodynamic points of supercritical Ar.
\pacs 65.20.De, 61.20.Lc, 66.10.cd
\end{abstract}


\section {Introduction}

Thermodynamics of the liquid state of matter was successfully
explained on the basis of thermodynamic perturbation theory (TDPT) more than
40 years ago  \cite{Han,Bar67,Bar76,Wee71}. Together with the integral equation
theory of static structure of liquids and computer simulations, it is the basis
of current exploration of physics and chemistry of real liquid systems.
Modern versions of TDPT  \cite{Mel09,Nez10}
were recently developed on the basis of treatment of the interaction interatomic
potential as a sum of an improved reference system
(hard-sphere- or soft-sphere-like repulsion plus short-range
Yukawa-like attraction) and a perturbation due to the remaining tail of the interparticle
interaction.

It is also possible to represent the thermodynamics of the liquid state
via time-dependent processes because
dynamic processes in condensed matter systems are tightly
connected with their thermodynamics. However, in contrast to solids,
where thermodynamics can be easily defined by contributions
from static lattice and small vibrations around the lattice sites
in terms of phonon excitations  \cite{Kit}, in fluids such an approach
is not applicable due to the dynamic disorder and the absence of stable
local energy minima for atomic particles. Furthermore, in liquid state
there exist various relaxation
processes which are absent in the solid state. As an example one can consider
a relaxation connected
with  diffusivity of local temperature. In liquid system without any external
temperature gradient, this relaxation
process is caused by local temperature gradients which
emerge due to {\it adiabatic} propagation of the most long-wavelength acoustic
excitations. The acoustic modes, which propagate with small wavenumbers
due to local conservation laws, in fact create adiabatically compressed
and decompressed local regions in liquid with different
local temperatures. Such a complex origin of collective
dynamics in fluids which is completely different from the elastic
sound propagation in solids is tightly connected with the
phenomenon of viscoelasticity of liquids  \cite{Boo}. Furthermore, the {\it acoustic
excitations}
in liquid state are only of longitudinal polarization (transverse
long-wavelength
acoustic modes are not supported in liquids, there can exist only short-wavelength
shear excitations with a propagation gap in the long-wavelength region)
and have non-zero damping. The dispersion law of longitudinal and
transverse collective excitations in liquids cannot be considered as
the linear in wave numbers  \cite{Bry10,Bry11b} as this is taken in the
Debye model of the heat capacity of solids. All these facts must
be reflected in the construction of liquid thermodynamics represented via
the dynamic processes.

Heat capacity being one of the important quantities in
thermodynamics of the liquid state of matter can be easily calculated
from the temperature fluctuations in the liquid system  \cite{Leb67}.
This is by date the most simple way of estimation of the specific
heat at constant volume $C_{\mathrm{v}}$ for realistic liquids from computer
simulations.
Schofield  \cite{Sch66,Sch68} has shown how the generalized wavenumber-dependent
thermodynamic quantities can be estimated from static correlators of different
statistically independent dynamic variables  \cite{Cop75}.
The generalized wavenumber-dependent thermodynamic quantities were studied later
on for pure  \cite{deS88,Mry95,Bry01} and binary liquids \cite{Bry97} in classical
and {\it ab initio}  \cite{Bry13b} molecular dynamics simulations.

Perhaps the first application of the hydrodynamic theory to collective dynamics
of liquids was reported in the paper by Landau and Placzek  \cite{Lan34}, while systematic
theoretical studies of liquid dynamics with hydrodynamic equations have started since
the seminal studies by Mountain  \cite{Mou66}. In 1971 Cohen et al. derived analytical
expressions for hydrodynamic time correlation functions for pure and binary liquids  \cite{Coh71},
which contained the so-called ``non-Lorentzian corrections'' and satisfied all the sum rules
existing within the hydrodynamic treatment \cite{Bha74}. Molecular dynamics (MD) simulations being
a perfect tool for exploration of time-dependent correlations in realistic liquids and solids
enable one to check the predictions of the hydrodynamic theory.

In this study we report the behaviour of autocorrelation functions of energy, heat
and entropy densities obtained by molecular dynamics simulations of supercritical Ar
at two densities. By date there were no MD studies of the entropy density autocorrelation
functions of pure liquids, for which the hydrodynamics predicts purely single-exponential
decay with time in the long-wavelength region. Besides, the zero-time value of the entropy
density and heat density autocorrelation functions should be connected with the
wavenumber-dependent specific heats $C_{\mathrm{p}}$ and $C_{\mathrm{v}}$, respectively. We will perform a
check of the long-wavelength limits of the $C_{\mathrm{p}}(k)$ and $C_{\mathrm{v}}(k)$ with the macroscopic
experimental values for these thermodynamic quantities  \cite{NIST}. The remaining part
of the paper is organised as follows. In the next section we give the predictions of the
hydrodynamic theory on the studied time correlation functions. In section~3 we supply
details of the
MD simulations, and section~4 reports obtained the results on the autocorrelation functions
of energy, heat and entropy densities obtained by molecular dynamics simulations of
su\-per\-cri\-ti\-cal Ar, their corresponding time-Fourier transforms, and the estimated
dependencies for the generalized specific heats $C_{\mathrm{p}}(k)$ and $C_{\mathrm{v}}(k)$. The last section
contains conclusions of this study.

\section {Hydrodynamic approach}

Hydrodynamic approach to collective dynamics of liquids is valid only on macroscopic
length and time scales where the most long-time relaxation processes in continuum
 are considered. The macroscopic equations in the hydrodynamic theory are in fact
the local conservation laws for the densities of particles, of total momentum and of
energy. For the case of longitudinal dynamics, the only relevant on macroscopic scale
contributions to the hydrodynamic time correlation functions  \cite{Boo} are the ones
coming from
acoustic longitudinal propagating modes with the linear in wave numbers $k$ dispersion law
\[
\omega_{\mathrm{hyd}}(k)=c_{\mathrm{s}}k
\]
and quadratic in $k$ damping
\[
\sigma_{\mathrm{hyd}}(k)=\Gamma k^2.
\]
Here $c_{\mathrm{s}}$ is the adiabatic speed of sound and the damping coefficient
$\Gamma=(D_{\mathrm{L}}+(\gamma-1)D_{\mathrm{T}})/2$ depends on the longitudinal kinematic viscosity $D_{\mathrm{L}}$,
the thermal diffusivity $D_{\mathrm{T}}$ and the ratio of specific heats $\gamma=C_{\mathrm{p}}/C_{\mathrm{v}}$.
The only relevant relaxing mode for the longitudinal dynamics on macroscopic scales is
the one coming from thermal relaxation, i.e., connected with the diffusivity of local
temperature $D_{\mathrm{T}}$.

The analytical expressions for the hydrodynamic time correlation functions, which
we will study by MD simulations, have in general a three term form: an exponential
function connected with thermal relaxing mode and two oscillating terms coming from
the acoustic excitations  \cite{Coh71}. The heat density autocorrelation function
reads
\begin{equation}\label{fhh_hyd}
F_{hh}(k,t)=A_{hh}\re^{-D_{\mathrm{T}}k^2t}+\left[B_{hh}\cos{c_{\mathrm{s}}kt}+D_{hh}(k)\sin{c_{\mathrm{s}}kt}\right]\re^{-\Gamma k^2t}~,
\end{equation}
where the amplitudes of mode contributions are constrained by the lowest-order sum rule
\begin{equation}\label{fhh_0}
A_{hh}(k)+B_{hh}(k)=k_{\mathrm{B}}T^2C_{\mathrm{v}}(k)~.
\end{equation}
Note that the mode contributions $A_{hh}$ and $B_{hh}$ are constants
in the long-wavelength limit, while the amplitide of the ``non-Lorentzian''
term $D_{hh}(k)$ linearly depends on $k$. However, outside the hydrodynamic
region  when one applies MD simulations the amplitudes $A_{hh}$ and
$B_{hh}$ become $k$-dependent making possible estimation of the wave-number
dependent specific heat. Similar is the expression for the energy density autocorrelation function:
\begin{equation}\label{fee_hyd}
F_{ee}(k,t)=A_{ee}\re^{-D_{\mathrm{T}}k^2t}+\left[B_{ee}\cos{c_{\mathrm{s}}kt}+D_{ee}(k)\sin{c_{\mathrm{s}}kt}\right]\re^{-\Gamma k^2t}~,
\end{equation}
while for the entropy density autocorrelation function, the hydrodynamics predicts a single
exponential form for pure liquids:
\begin{equation}\label{fss_hyd}
F_{ss}(k,t)=A_{ss}\re^{-D_{\mathrm{T}}k^2t}~,
\end{equation}
with
\begin{equation}\label{fss_0}
A_{ss}(k)=k_{\mathrm{B}}C_{\mathrm{p}}(k)~.
\end{equation}
Hence, the wavenumber-dependent specific heats $C_{\mathrm{v}}(k)$ and $C_{\mathrm{p}}(k)$ are connected
with the static heat-density and entropy-density autocorrelation functions, which
in fact are the zeroth moments of the corresponding spectral functions $S_{hh}(k,\omega)$
and $S_{ss}(k,\omega)$. In the long-wavelength limit, the wavenumber dependent
specific heats tend to their macroscopic values $C_{\mathrm{v}}$ and $C_{\mathrm{p}}$. The advantage of the
hydrodynamic approach lies in a possibility to estimate the contributions from relaxing
and propagating modes to corresponding thermodynamic quantities  \cite{Bry13}.

\section {Molecular dynamics simulations}

We performed molecular dynamics simulations for supercritical Ar at
temperature $T=280$~K and two densities 750~kg/m$^3$ and 1464.5~kg/m$^3$
using a system of 2000 particles interacting via the Woon potential
 \cite{Woo93} with the parameters taken from  \cite{Bom00}. The cut-off
radius for the effective two-body potential was
12~\AA. This potential permits a nice reproduction of the thermodynamic and
dynamic quantities of supercritical Ar by MD simulations as it was shown in
 \cite{Bry11}.

The simulations were performed in the microcanonical ensemble
with the time step of 2~fs. The duration of production runs was of 360 000
time steps.
Every sixth configuration was used for the sampling of dynamic
variables of particle density, momentum density, energy density and first time
derivative of the momentum density.
The averages of static and time
correlation functions over all possible directions of different
wave vectors with the same wave number were performed.

During the MD simulations we sampled the spatial Fourier components
of hydrodynamic variables of: particle density
\begin{equation} \label{nkt}
n(k,t)=\frac{1}{\sqrt{N}}\sum_{i=1}^N
\re^{-\ri {\bf k}{\bf r}_i(t)}~,
\end{equation}
density of longitudinal momentum
\begin{equation} \label{jlkt}
J^{\mathrm{L}}(k,t)=\frac{m}{k\sqrt{N}}\sum_{i=1}^N ({\bf k}\bf{v}_i)
\re^{-\ri {\bf k}{\bf r}_i(t)}~,
\end{equation}
and energy density
\begin{equation} \label{ekt}
e(k,t)=\frac{1}{\sqrt{N}}\sum_{i=1}^N
\varepsilon_i(t)\re^{-\ri {\bf k}{\bf r}_i(t)}~,
\end{equation}
where ${\bf r}_i(t)$, ${\bf v}_i(t)$ and
$\varepsilon_i(t)$ are the trajectories, velocities
and single-particle energies of the $i$-th particle  \cite{Cop75}.
The single-particle energies were calculated as follows:
\[
\varepsilon_i(t)=\frac{mv_i^2(t)}{2}+\frac{1}{2}\sum_{j\neq i}\Phi
\left(|{\bf r}_i(t)-{\bf r}_j(t)|\right),
\]
where $\Phi(r)$ is the effective pair potential used in the simulations.

The above dynamic variables were used to calculate the hydrodynamic
time correlation functions as well as the heat density
autocorrelation functions
\begin{equation} \label{fhh_def}
F_{hh}(k,t)=\left\langle h(k,t)h^*(k,t=0)\right\rangle
\end{equation}
with
\begin{equation} \label{hkt}
h(k,t)=e(k,t)-\frac{f_{ne}}{f_{nn}}n(k,t),
\end{equation}
where the brackets mean the ensemble average and static
averages read as follows $f_{ne}(k)=\langle n(k)e(-k)\rangle$
and $f_{nn}(k)=\langle n(k)n(-k)\rangle$.

In order to study the entropy density autocorrelation functions
$F_{ss}(k,t)$, we sampled the density of the longitudinal
component of stress tensor $\sigma^{\mathrm{L}}(k,t)$ via  \cite{Cop75}
\[
{\dot{J}}^{\mathrm{L}}(k,t)=-\ri k\sigma^{\mathrm{L}}(k,t)~,
\]
where the overdot denotes the time derivative of the corresponding
dynamic variable. Hence, the dynamic variable of entropy density
can be sampled from MD simulations as  \cite{Cop75}
\[
s(k,t)=\frac{1}{T}\left[e(k,t)-\frac{f_{e\sigma^{\mathrm{L}}}}{f_{\sigma^{\mathrm{L}}n}}n(k,t)\right]
\]
with the corresponding static averages
$f_{e\sigma^{\mathrm{L}}}(k)=\langle e(k)\sigma^{\mathrm{L}}(-k)\rangle$
and $f_{\sigma^{\mathrm{L}}n}(k)=\langle \sigma^{\mathrm{L}}(k)n(-k)\rangle$.

\section {Results and discussion}

\begin{figure}[!b]
\centerline{
\includegraphics[width=0.5\textwidth]{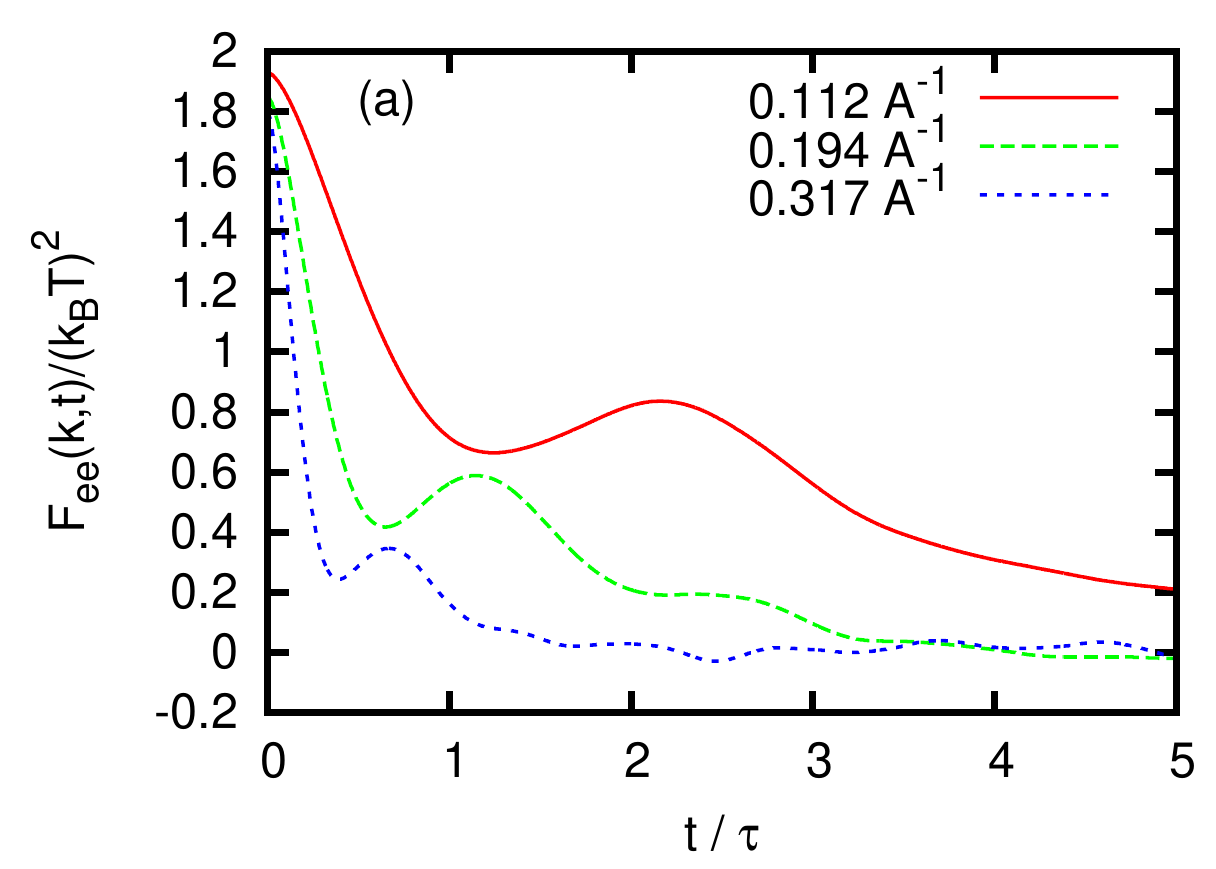}
\includegraphics[width=0.5\textwidth]{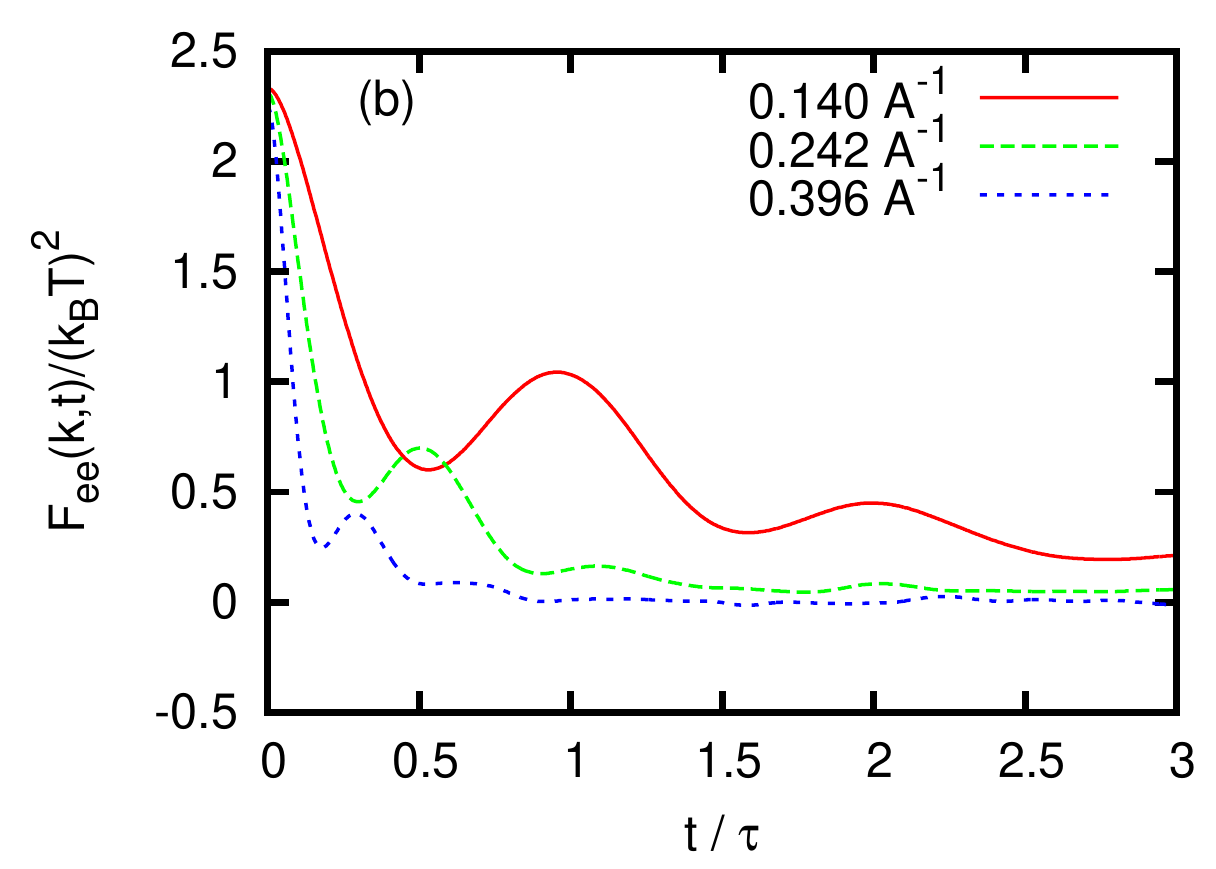}
}
\caption{(Color online) Energy density autocorrelation functions $F_{ee}(k,t)$
at three wave numbers as calculated from MD simulations for
supercritical Ar at $T=280$~K and densities 750~kg/m$^3$ (a) and
1464.5~kg/m$^3$ (b).
The time scales $\tau$ are equal to 3.701~ps (a) and  2.961~ps (b).
} \label{fee}
\end{figure}
\begin{figure}[htb]
\centerline{
\includegraphics[width=0.5\textwidth]{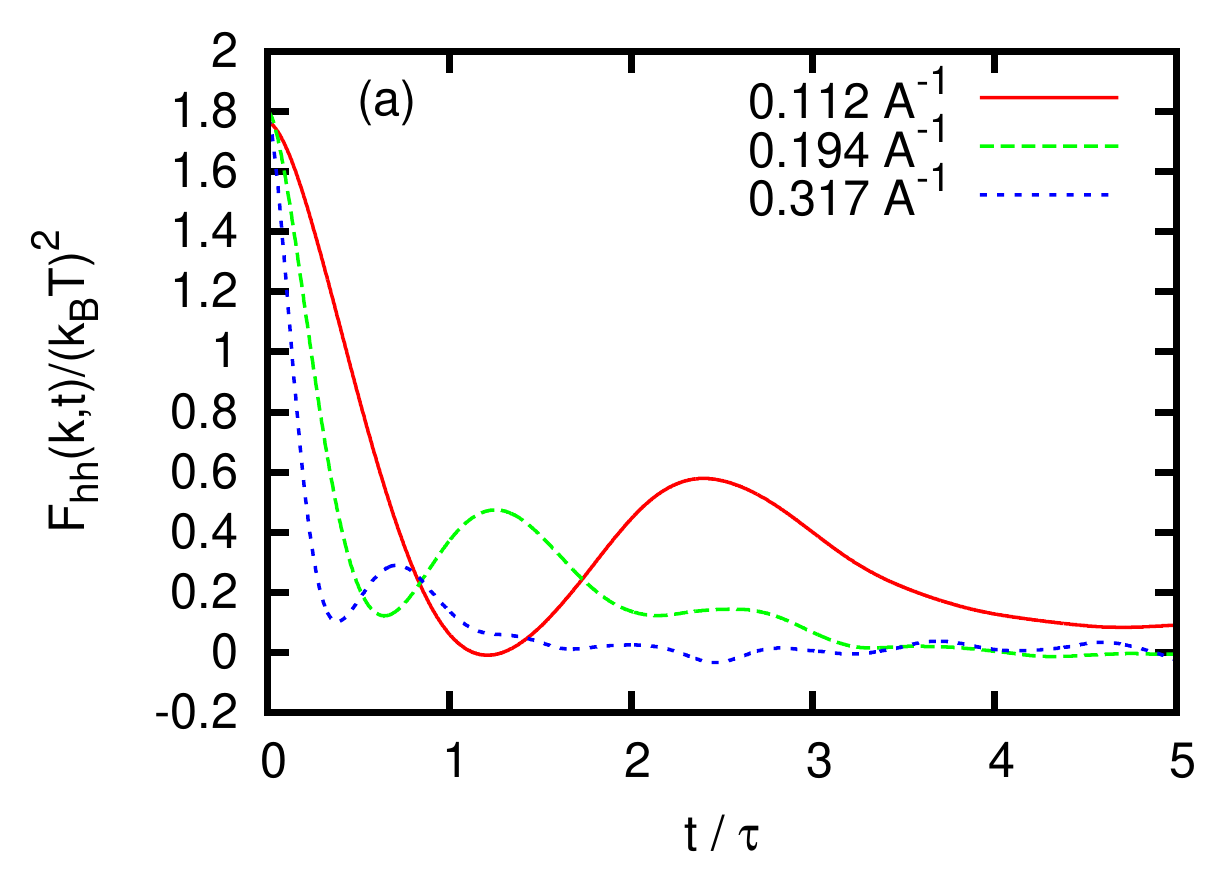}
\includegraphics[width=0.5\textwidth]{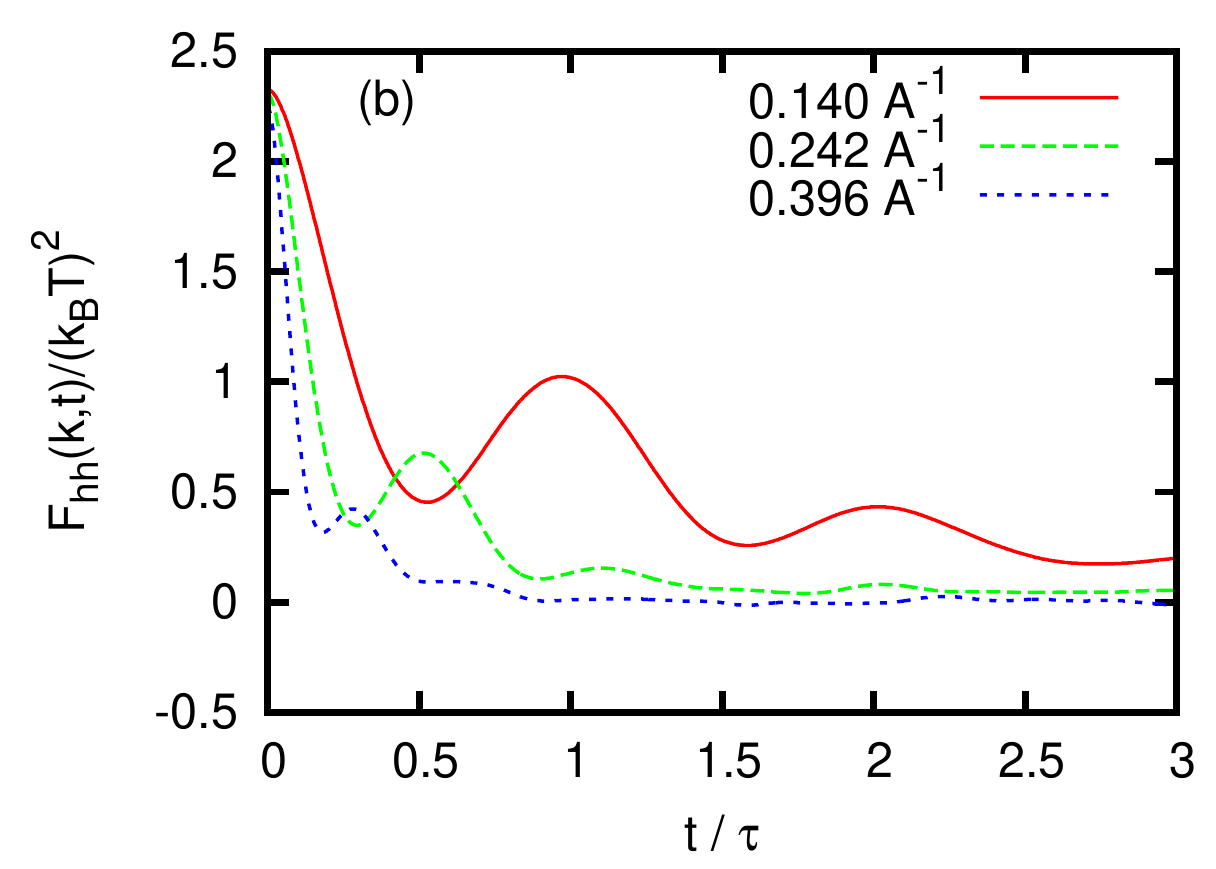}
}
\caption{(Color online) Heat density autocorrelation functions $F_{hh}(k,t)$
at three wave numbers as calculated from MD simulations for
supercritical Ar at $T=280$~K and densities 750~kg/m$^3$ (a) and
1464.5~kg/m$^3$ (b).
The time scales $\tau$ are equal to 3.701~ps (a) and  2.961~ps (b).
} \label{fhh}
\end{figure}

Energy density autocorrelation functions $F_{ee}(k,t)=\langle e(k,t)\re^*(k,t=0)\rangle$
calculated in MD simulations are shown for three wavenumbers in small-$k$ region in
figure~\ref{fee}. The time dependence of these time correlation functions is very similar
to the relaxation of the heat density autocorrelations shown in figure~\ref{fhh}. A typical
feature is an increasing amplitude and a decreasing frequency of the damped oscillations
in the shape of $F_{ee}(k,t)$ and $F_{hh}(k,t)$ towards smaller wave numbers, which is
typical of the contribution coming from acoustic excitations. The
relative strength of the relaxing and oscillating contributions in $F_{hh}(k,t)$ was
estimated in the long-wavelength limit in  \cite{Bry13}  as the inverse of $(\gamma -1)$,
i.e., the inverse of the Landau-Placzek ratio known for the density-density time correlation
functions. Hence, for the case of the two thermodynamic points of the supercritical Ar
studied here and having the experimental ratio of specific heats $\gamma$ of 2.30 and 1.55
 \cite{NIST} for the
low- and high-density studied systems, one should observe much smaller relaxing contribution
for the low-density state.
Indeed, in figure~\ref{fhh}~(a) the oscillations in the shape of $F_{hh}(k,t)$ even
make the heat density autocorrelation function negative  close to $\omega\tau\sim 1$ for the
smallest possible in these simulations wave number, while in figure~\ref{fhh}~(b), due to the much stronger
contribution from the relaxing thermal mode, the $F_{hh}(k,t)$ is positive for all times.
It is important that the dimensionless functions $F_{hh}(k,t)$ have the zero-time values
tending to the macroscopic values of $C_{\mathrm{v}}$ due to (\ref{fhh_0}). Indeed, in figure~\ref{fhh}~(a) the
zero-time values tend to the macroscopic value of $C_{\mathrm{v}}=1.775$  \cite{NIST}, while for the high-density
state they tend to the value $C_{\mathrm{v}}=2.34$.
\begin{figure}[!h]
\centerline{
\includegraphics[width=0.5\textwidth]{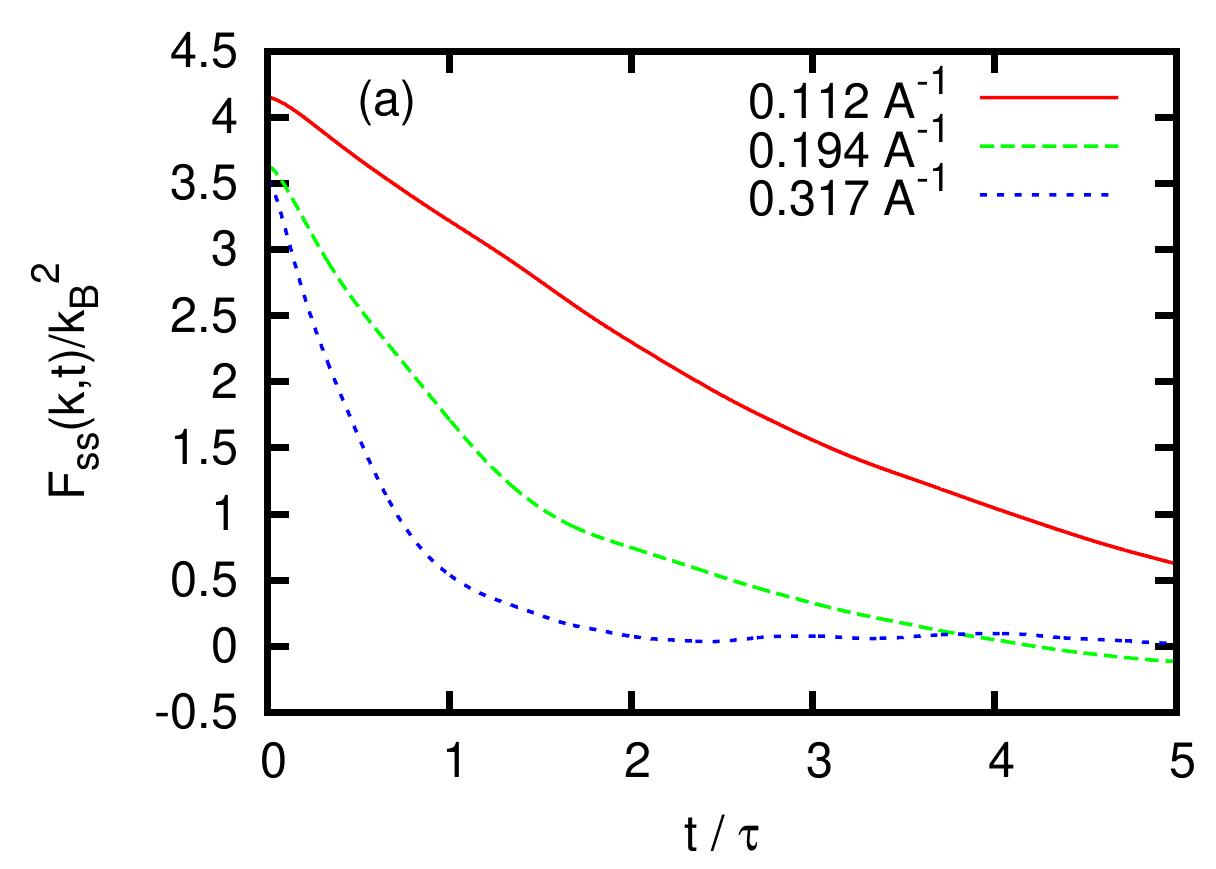}
\includegraphics[width=0.5\textwidth]{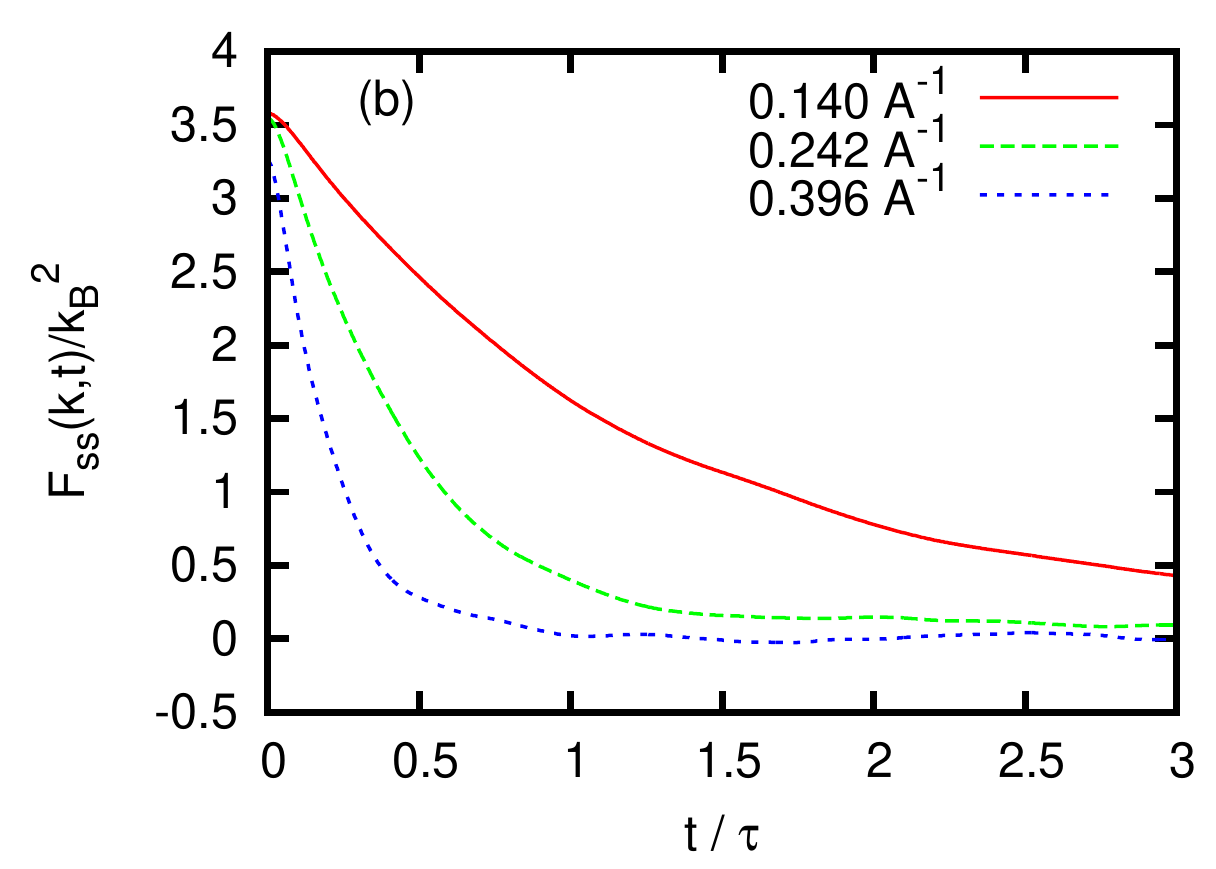}
}
\caption{(Color online) Entropy density autocorrelation functions $F_{ss}(k,t)$
at three wave numbers as calculated from MD simulations for
supercritical Ar at $T=280$~K and densities 750~kg/m$^3$ (a) and
1464.5~kg/m$^3$ (b).
The time scales $\tau$ are equal to 3.701~ps (a) and  2.961~ps (b).
} \label{fss}
\end{figure}

The entropy density autocorrelation functions $F_{ss}(k,t)$ calculated
in MD simulations for two densities of the supercritical Ar (figure~\ref{fss})
show for the smallest wave numbers a simple relaxation shape without any
oscillating contribution. This is in complete agreement with the
prediction of the hydrodynamic theory (\ref{fss_hyd}). The zero-time values
of $F_{ss}(k,t)$ (\ref{fss_0}) should tend in the long-wavelength limit to the macroscopic
value of $C_{\mathrm{p}}$. Indeed, one observes in figures~\ref{fss}~(a) and (b) that in agreement
with this hydrodynamic prediction, the zero-time values of the normalized
functions tend to the macroscopic values of $C_{\mathrm{p}}$ reported in the
NIST database: 4.097 for the low-density state and 3.62 for the high-density state.

\begin{figure}[htb]
\centerline{
\includegraphics[width=0.5\textwidth]{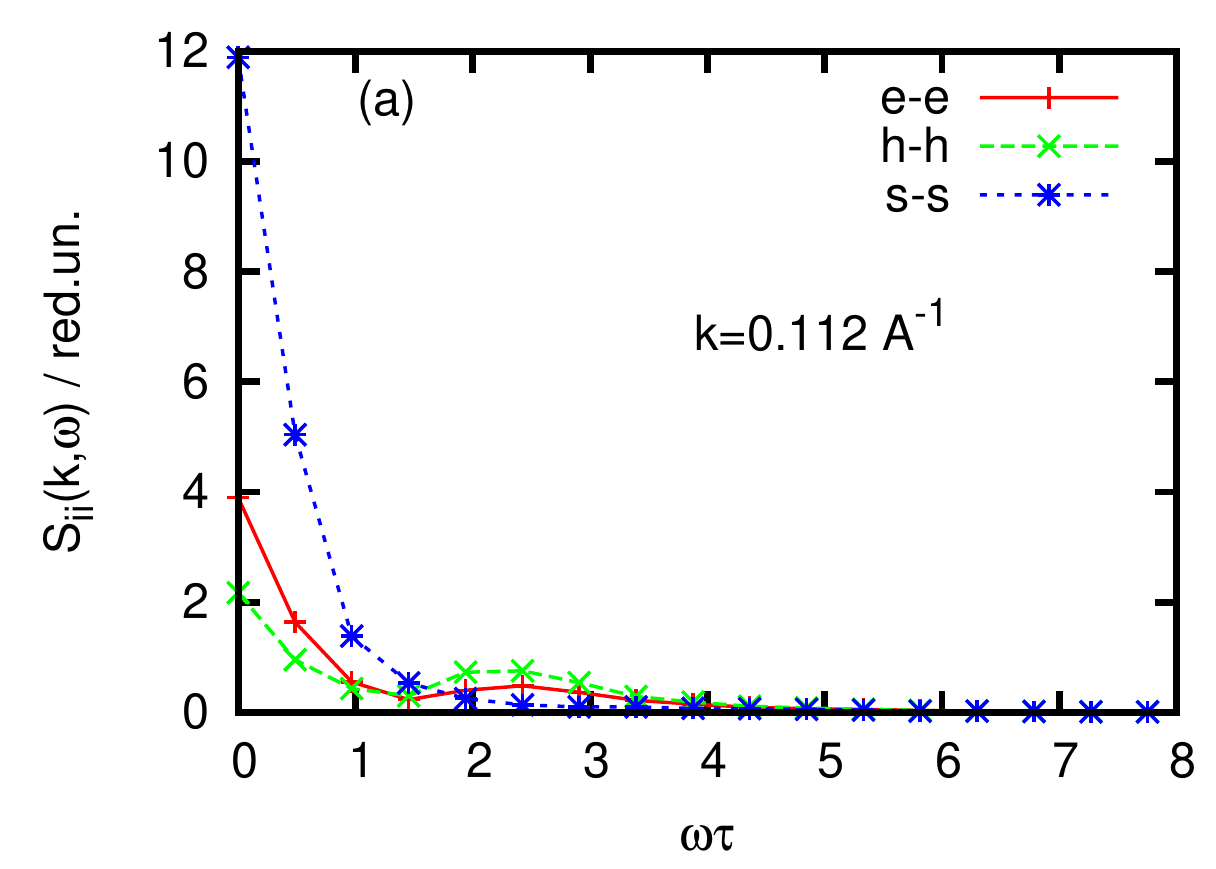}
\includegraphics[width=0.5\textwidth]{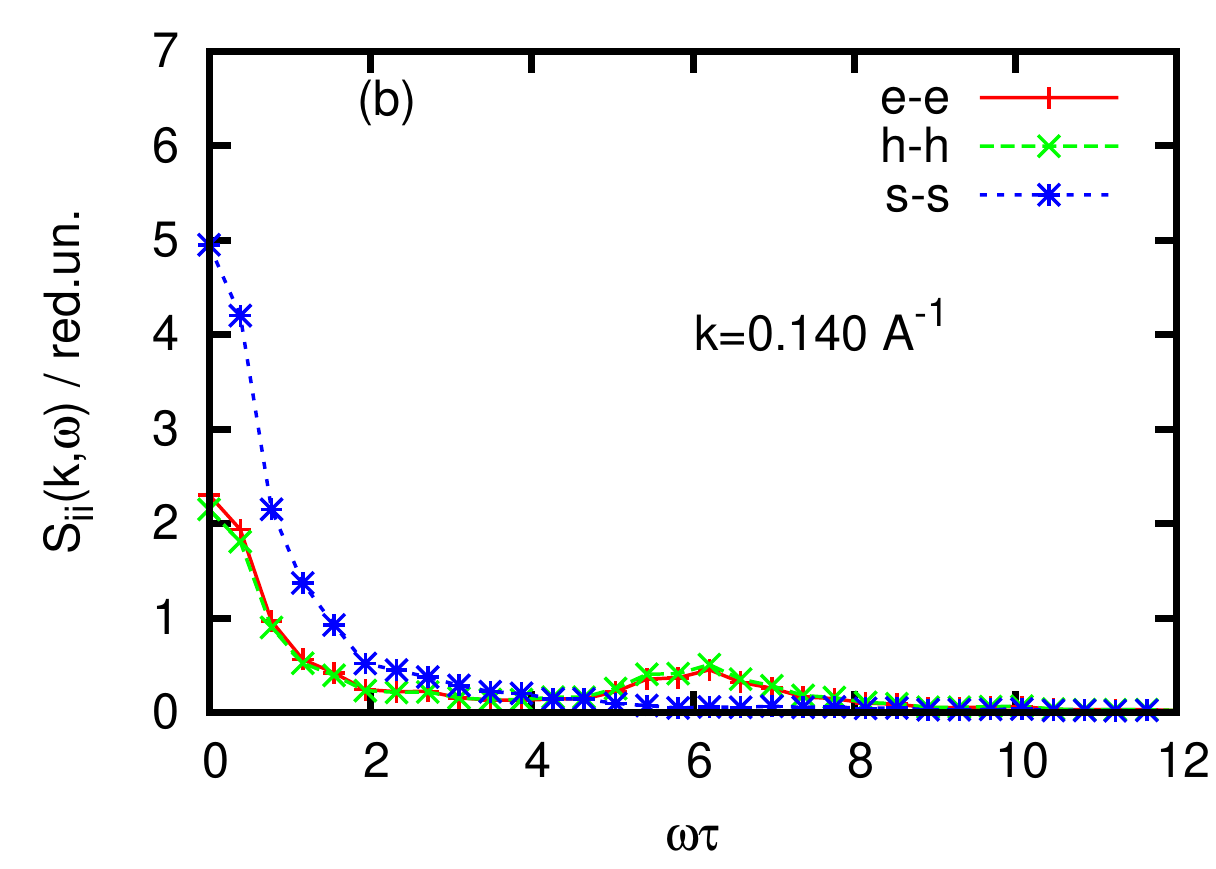}
}
\caption{(Color online) Spectral functions energy-energy $S_{ee}(k,\omega)$,
heat-heat $S_{hh}(k,\omega)$, entropy-entropy $S_{ss}(k,\omega)$
at the smallest available in MD simulations wave numbers for
supercritical Ar at $T=280$~K and densities 750~kg/m$^3$ (a) and
1464.5~kg/m$^3$ (b).
The time scales $\tau$ are equal to 3.701~ps (a) and  2.961~ps (b).
} \label{sii}
\end{figure}

The time-Fourier transformed energy density (e-e), heat energy density (h-h)
and entropy density (s-s) autocorrelation functions, $S_{ii}(k,\omega),~i=e,h,s$
are shown for the two studied densities of the supercritical Ar in figure~\ref{sii}.
The $S_{ee}(k,\omega)$ and $S_{hh}(k,\omega)$ have the typical of the dynamic
structure factors shape with the central peak due to thermal relaxation and
the side peak due to propagating acoustic excitations. The ratio of the integral
intensities of the central and two side peaks known as the Landau-Placzek-type ratio
was obtained in  \cite{Bry13} for the case of $S_{hh}(k,\omega)$ and is equal to
$(\gamma-1)^{-1}$. The side peak of the $S_{ee}(k,\omega)$ and $S_{hh}(k,\omega)$
is the consequence of the corresponding oscillating contributions to the $F_{ee}(k,t)$
and $F_{hh}(k,t)$, respectively. It is remarkable that the calculated spectral
function $S_{ss}(k,\omega)$ shows only the one-peak shape with the central peak
due to thermal relaxation (line connected star symbols in figure~\ref{sii}).
\begin{figure}[htb]
\centerline{
\includegraphics[width=0.5\textwidth]{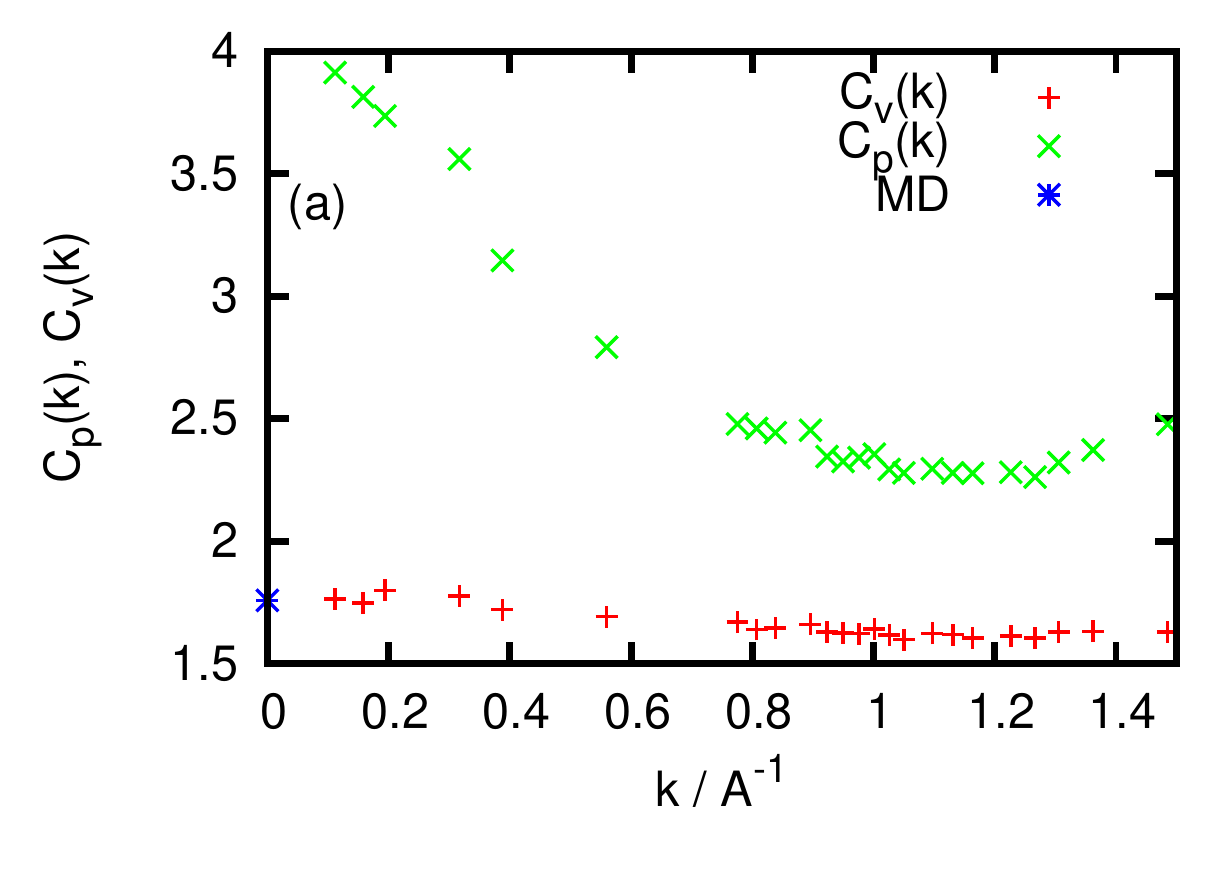}
\includegraphics[width=0.5\textwidth]{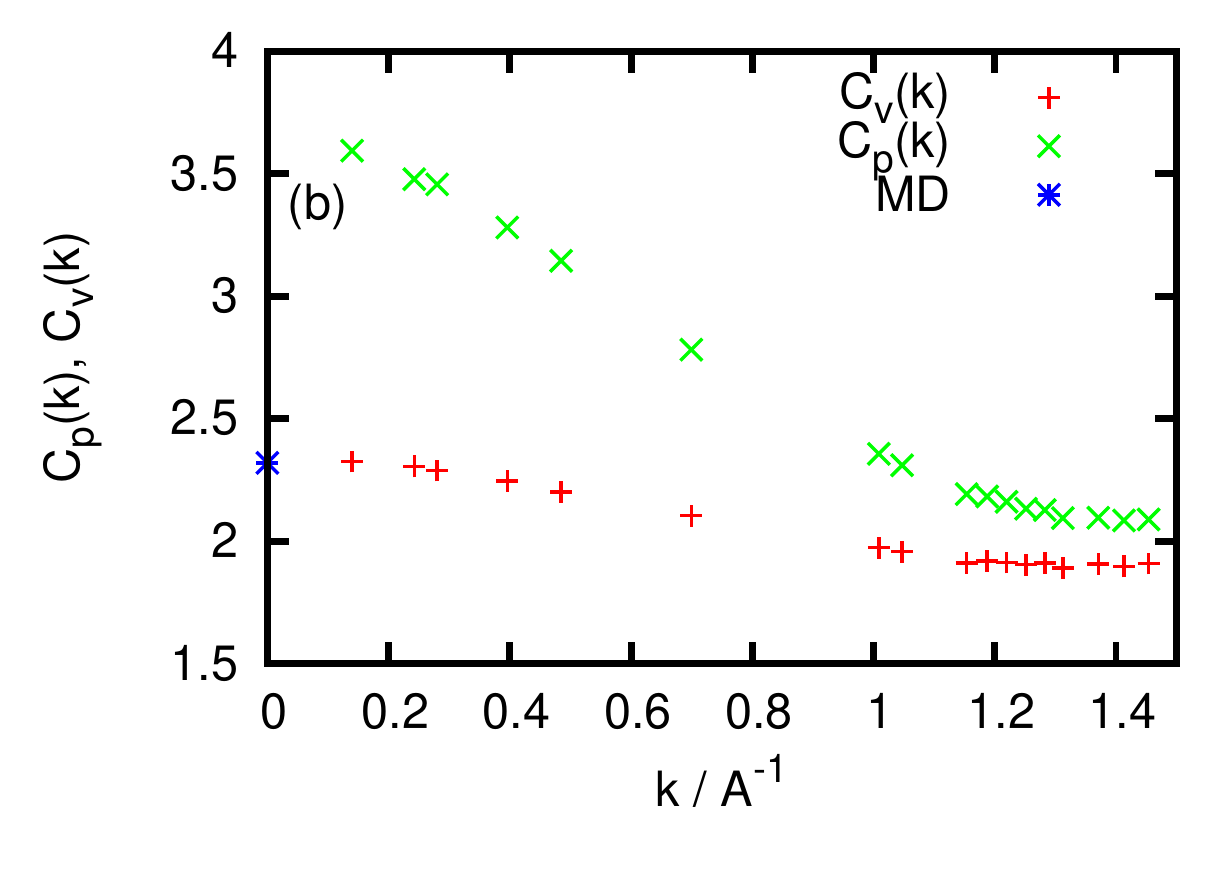}
}
\caption{(Color online) Generalized wavenumber-dependent specific heats at
constant volume $C_{\mathrm{v}}(k)$ and constant pressure $C_{\mathrm{p}}(k)$ for
supercritical Ar at $T=280$~K ad densities 750~kg/m$^3$ (a) and
1464.5~kg/m$^3$ (b). The asterisks at $k=0$ correspond to
macroscopic values of $C_{\mathrm{v}}$ calculated from the temperature
fluctuations  \cite{Leb67}.
} \label{cvcp}
\end{figure}

The zero-time values of the time correlation functions $F_{hh}(k,t)$
and $F_{ss}(k,t)$ permit estimation of the wavenumber-dependent
specific heats $C_{\mathrm{v}}(k)$ and $C_{\mathrm{p}}(k)$, which in the long-wavelength
limit tend to their macroscopic values. In figure~\ref{cvcp} we show
the wavenumber dependencies of $C_{\mathrm{v}}(k)$ and $C_{\mathrm{p}}(k)$ for the two
densities of the supercritical Ar at temperature 280~K. The asterisks
at $k=0$ show the values of macroscopic values of $C_{\mathrm{v}}$ obtained
from the standard expression via the thermal fluctuations in the
fluid system  \cite{Leb67}. One can see that the reported wave-number
dependencies tend quite well to the experimental values for the
supercritical Ar taken from the NIST database  \cite{NIST}:
$C_{\mathrm{v}}$=1.775$k_{\mathrm{B}}$ and $C_{\mathrm{p}}$=4.097$k_{\mathrm{B}}$ for the density of 750~kg/m$^3$
and $C_{\mathrm{v}}$=2.34$k_{\mathrm{B}}$ and $C_{\mathrm{p}}$=3.62$k_{\mathrm{B}}$ for the density of 1464.5~kg/m$^3$.

\section {Conclusions}

In the MD simulations we calculated the autocorrelation functions of
energy density, heat density and energy density and compared them for
the smallest available wave numbers with the predictions of the hydrodynamic
theory. Our results give evidence of a nice correspondence of the hydrodynamic
description of time correlation functions and their zero-time values, which
made it possible to estimate the wavenumber-dependent specific heats $C_{\mathrm{v}}(k)$ and
$C_{\mathrm{p}}(k)$,  with the MD-based estimation of their macroscopic values and
experimental values for specific heats of supercritical Ar taken from the
NIST database.

These results are very important from the viewpoint of analysing the contributions
from thermal relaxation and collective excitations to the specific heats of
liquids. Our results and the hydrodynamic approach make evidence of the definitive
role of the thermal relaxation in calculation of the specific heats of fluids when
it is estimated via dynamic processes. These results will urge a further understanding
of the dynamics of the supercritical state of matter, which shows many interesting
features observed recently in the scattering experiments and MD simulations on supercritical Argon \cite{Sim10,Gor13} and Oxygen \cite{Gor06}.


%
\ukrainianpart

\title{Теплоємність рідин: Гідродинамічний підхід}
\author{Т.~Брик\refaddr{label1,label2}, Т.~Скопіньйо\refaddr{label3},
Дж.~Руокко\refaddr{label3,label4}}

\addresses{
\addr{label1} Інститут фізики конденсованих систем НАН України,
79011 Львів, Україна
\addr{label2} Інститут прикладної математики та фундаментальних наук,
Національний Університет ``Львівська Політехніка'', 79013 Львів, Україна
\addr{label3} Фізичний факультет, Римський університет ``La Sapienza'',
I--00185, Рим, Італія
\addr{label4} Центр біо-нано наук Sapienza, Італійський технологічний
інститут, 295 вул. королеви Елени, I--00161, Рим, Італія
}

\makeukrtitle

\begin{abstract}
\tolerance=3000%
Ми досліджуємо автокореляційні функції густин енергії, тепла та ентропії,
обчислених при моделюванні методом молекулярної динаміки для надкритичного
Ar та порівнюємо їх із передбаченнями гідродинамічної теорії.
Показано, що передбачена гідродинамічною теорією одноекспонентна форма
автокореляційної функції густини ентропії чудово відтворюється при малих
хвильових числах молекулярною динамікою та дозволяє розрахунок залежної
від хвильового числа питомої теплоємності при сталому тиску.
Отримані питомі теплоємності, залежні від хвильового числа,
при сталому об'ємі та тиску, $C_v(k)$ and $C_p(k)$,
у довгохвильовій границі добре узгоджуються з макроскопічними експериментальними значеннями $C_v$ and $C_p$
для досліджених термодинамічних точок надкритичного Ar.

\keywords рідини, термодинаміка, теплоємність, гідродинамічна теорія

\end{abstract}


\begin{thebibliography}{12}
\bibitem{Han}   Hansen~J.-P., McDonald~I.R., Theory of Simple Liquids, London, Academic, 1986.
\bibitem{Bar67} Barker~J.A., Henderson~D., J. Chem. Phys., 1967, \textbf{47}, 2856;
                \bibdoi{10.1063/1.1712308}.
\bibitem{Bar76} Barker~J.A., Henderson~D.,  Rev. Mod. Phys., 1976, \textbf{48}, 587;
                \bibdoi{10.1103/RevModPhys.48.587}.
\bibitem{Wee71} Weeks~J.D., Chandler~D., Andersen~H.C., J. Chem. Phys., 1971, \textbf{54}, 5237;
                \bibdoi{10.1063/1.1674820}.
\bibitem{Mel09} Melnyk~R., Nezbeda~I., Henderson~D., Trokhymchuk~A.,
                Fluid Phase Equilibr., 2009, \textbf{279}, 1;\\
                \bibdoi{10.1016/j.fluid.2008.12.004}.
\bibitem{Nez10} Nezbeda~I.,  Melnyk~R., Trokhymchuk~A., J. Supercritical
                Fluid, 2010, \textbf{55}, 448;
                \bibdoi{10.1016/j.supflu.2010.10.041}.
\bibitem{Kit}   Kittel~Ch., Introduction to Solid State Physics, 8-th Edition,
                New York, Wiley, 2004.
\bibitem{Boo}   Boon~J.-P., Yip~S., Molecular Hydrodynamics, New-York, McGraw-Hill, 1980.
\bibitem{Bry10} Bryk~T., Mryglod~I., Scopigno~T., Ruocco~G., Gorelli F.,
                Santoro~M., J. Chem. Phys., 2010, \textbf{133}, 024502;
                \bibdoi{10.1063/1.3442412}.
\bibitem{Bry11b} Bryk~T., Eur. Phys. J. Spec. Top., 2011, \textbf{196}, 65;
                \bibdoi{10.1140/epjst/e2011-01419-x}.
\bibitem{Leb67} Lebowitz~J.L., Perkus~J.K., Verlet~L., Phys. Rev., 1967, \textbf{153}, 250;
                \bibdoi{10.1103/PhysRev.153.250}.
\bibitem{Sch66} Schofield~P., Proc. Phys. Soc., 1966, \textbf{88}, 149;
                \bibdoi{10.1088/0370-1328/88/1/318}.
\bibitem{Sch68} Schofield~P., In: Physics of simple liquids, Temperley~H.N.V.,
                Rowlinson~J.S., Rushbrooke~G.S. (Eds.), North-Holland Publishing,
                Amsterdam, 1968.
\bibitem{Cop75} Copley~J.R.D., Lovesey~S.W., Rep. Prog. Phys., 1975, \textbf{38}, 461;
                \bibdoi{10.1088/0034-4885/38/4/001}.
\bibitem{deS88} de~Schepper~I.M., Cohen~E.G.D., Bruin~C., van~Rijs~J.C.,
                Montfrooij~W., de~Graaf~L.A., Phys. Rev. A, 1988, \textbf{38},
                271; \bibdoi{10.1103/PhysRevA.38.271}.
\bibitem{Mry95} Mryglod~I.M., Omelyan~I.P., Tokarchuk~M.V., Mol. Phys., 1995,
                \textbf{84}, 235; \bibdoi{10.1080/00268979500100181}.
\bibitem{Bry01} Bryk~T., Mryglod~I., Phys. Rev. E, 2001, \textbf{63}, 051202;
                \bibdoi{10.1103/PhysRevE.63.051202}.
\bibitem{Bry97} Bryk~T., Mryglod~I., Kahl~G., Phys. Rev. E, 1997, \textbf{56}, 2903;
                \bibdoi{10.1103/PhysRevE.56.2903}.
\bibitem{Bry13b} Bryk~T., Ruocco~G., Mol. Phys., 2013, \textbf{111}, 3457;
                \bibdoi{10.1080/00268976.2013.838313}.
\bibitem{Lan34} Landau~L.D., Placzek~G., Physik. Z. Sowjetunion., 1934,
                \textbf{5}, 172.
\bibitem{Mou66} Mountain~R.D., Rev. Mod. Phys., 1966,  \textbf{38}, 205;
                \bibdoi{10.1103/RevModPhys.38.205}.
\bibitem{Coh71} Cohen~C., Sutherland~J.W.H., Deutch~J.M.,
                Phys. Chem. Liq., 1971, \textbf{2}, 213;\bibdoi{10.1080/00319107108083815}.
\bibitem{Bha74} Bhatia~A.B., Thornton~D.E., March~N.H.,
                Phys. Chem. Liq., 1974, \textbf{4}, 97;\bibdoi{10.1080/00319107408084276}.
\bibitem{NIST} Lemmon~E.W., McLinden~M.O., Friend~D.G.,
                Thermophysical Properties of Fluid Systems, In:
               {NIST} Chemistry WebBook, {NIST} Standard
                Reference Database 69 (National Institute of Standards and Technology,
                Gaithersburg MD, 2004).\url{http://webbook.nist.gov}.

\bibitem{Bry13} Bryk~T., Ruocco~G., Scopigno~T., J. Chem. Phys., 2013, \textbf{138}, 034502;
                \bibdoi{10.1063/1.4774406}.
\bibitem{Woo93} Woon~D.E., Chem.Phys.Lett., 1993, \textbf{204}, 29;
                \bibdoi{10.1016/0009-2614(93)85601-J}.
\bibitem{Bom00} Bomont~J.-M., Bretonnet~J.-L., Pfleiderer~T., Bertagnolli~H.,
                J. Chem. Phys., 2000, \textbf{113}, 6815;
                \bibdoi{10.1063/1.1290131}.
\bibitem{Bry11} Bryk~T., Ruocco~G., Mol. Phys., 2011, \textbf{109}, 2929;
                \bibdoi{10.1080/00268976.2011.617321}.
\bibitem{Sim10} Simeoni~G., Bryk~T., Gorelli~F.A., Krisch~M., Ruocco~G.,
               Santoro~M., Scopigno~T., Nat. Phys., 2010, \textbf{6}, 503;
                \bibdoi{10.1038/nphys1683}.
\bibitem{Gor13} Gorelli~F.A., Bryk~T., Krisch~M., Ruocco~G.,
                Santoro~M., Scopigno~T., Sci. Rep., 2013, \textbf{3}, 01203;
                \bibdoi{10.1038/srep01203}.
\bibitem{Gor06} Gorelli~F.A., Santoro~M., Scopigno~T., Krisch~M., Ruocco~G.,
                Phys. Rev. Lett., 2006, \textbf{97}, 245702;
                \bibdoi{10.1103/PhysRevLett.97.245702}.
%
\end{thebibliography}
\end{document}